\documentclass[a4paper,12pt]{article}
\usepackage[english] {babel}
\usepackage{amsmath,amsfonts,amssymb,hyperref,graphicx,booktabs,cite}
\usepackage{bookmark}
\usepackage{authblk,appendix}
\makeatletter
\def\@maketitle{%
  \newpage
  \null
  \vskip 2em%
  \begin{center}%
  \let \footnote \thanks
    {\Large\bfseries \@title \par}%
    \vskip 1.5em%
    {\normalsize
      \lineskip .5em%
      \begin{tabular}[t]{c}%
        \@author
      \end{tabular}\par}%
    \vskip 1em%
    {\normalsize \@date}%
  \end{center}%
  \par
  \vskip 1.5em}
\makeatother

\renewcommand\appendix{\par
  \setcounter{section}{0}
  \setcounter{subsection}{0}
  \setcounter{figure}{0}
  \setcounter{table}{0}
  \setcounter{equation}{0}
  \renewcommand\thesection{Appendix \Alph{section}}
  \renewcommand\thesubsection{\Alph{section}.\arabic{subsection}}
  \renewcommand\thefigure{\Alph{section}\arabic{figure}}
  \renewcommand\thetable{\Alph{section}\arabic{table}}
   \renewcommand\theequation{\Alph{section}\arabic{equation}}
}
\hypersetup{
  pdfauthor={S.S. Baturin}
}

\begin{document}
\title{Calculation of the Cherenkov fields in the cross-section of a short relativistic bunch.}
\author{S.S. Baturin$^{1,2,}$%
  \thanks{Electronic address: \texttt{s.s.baturin@gmail.com}}}
\affil{1. St.Petersburg Electrotechnical University LETI, \\ St.Petersburg, Russia, 197376}
\affil{2. Euclid Techlabs, LLC, \\ Solon, Ohio, 44139, USA}
\date{Dated: \today}
\maketitle
\begin{abstract}
Recently, a new method for calculating the Cherenkov field acting on a point-like electron bunch passing through longitudinally homogeneous structures lined with arbitrary slowdown layers was proposed, where the formalism was obtained though consideration  of a general integral relation that allows calculation of the fields at the vicinity of a point-like bunch. It demonstrates that the Cherenkov field at the point of the short relativistic bunch does not depend on the waveguide system material and is a constant for any given transverse dimensions and cross-section shapes of waveguides. With this paper we present a strict derivation of the fields formulas valid at the cross-section of a bunch on the basis of a conformal mapping method. We generalize the results of the previous paper to the case of transversely distributed bunch by deriving a two-dimensional Green function at the cross-section of a bunch. Comparison of the results proving validity of the method is given in the appendix.  
\end{abstract}

\pdfbookmark[section]{\contentsname}{toc}
\section{Introduction}
Relativistic, high intensity and small emittance electron bunches are the basis of linear collider (ILC \cite{ILC}, CLIC \cite{CLIC}) and FEL (LCLS, X-FEL, etc.) \cite{Bost} projects among many others. These bunches excite Cherenkov wakefields when electrons pass through the accelerating structures or other longitudinally extended components of a beam line (pipes, collimators, bellows). Theoretical analysis of Cherekov radiation commonly considers a "short bunch" approach \cite{Bane1,Bane2,Novo,Bane3}. It can be applied for various Cherenkov generation parameters, where the moving bunch size is much less that the fundamental wavelength if the high frequency spectrum range is not under investigation. 
Recently a new theoretical approach that can be used for obtaining direct analytical formulas for electromagnetic field components at the position of a point-like bunch was proposed \cite{myPRL,myArxiv,myArxiv_leont}. It was demonstrated that the longitudinal part of the Lorentz force acting on the point-like bunch does not depend on the waveguide system material and is a constant for any given transverse dimensions and the cross-section shapes of waveguides. The equivalence and exact matching of the longitudinal electric fields $E_z$ on beams passing through various waveguide configurations was also subject to analysis. The proposed approach considers using an integral relation based on the cylindrical slow-wave structure model. For planar, square and other cross-section geometries, one can obtain a corresponding form factor multiplier by using a conformal mapping from the solution for the cylindrical case \cite{myPRL,myArxiv,myArxiv_leont}. 
However, the question about the transverse distributions of the fields and Lorentz force in the cross-section of a bunch was not examined in details.  The method described in \cite{myPRL,myArxiv} gives the opportunity to calculate the longitudinal electric field $E_z$ and longitudinal derivative of the transverse part of the Lorentz force at the point of a bunch, which is also assumed to be point-like. In this paper we are going to extend the idea of \cite{myPRL,myArxiv} to a bunch which is a point only by longitudinal coordinate and has some arbitrary distribution by transverse coordinates and derive formulas that gives distribution by the transverse coordinates of the longitudinal fields $E_z$, $H_z$ and the transverse part of the Lorentz force in the cross-section of a bunch. This formulas are the Green function and therefore may be applied to the fields calculation of distributed bunches by transverse coordinates.
We also show in Section \ref{sec:GT} how the relativistic Gauss theorem \cite{myPRL} can be derived with the use of the proposed approach.   

\section{Main formular derivation}\label{sec:PF}
In this section, we are going to derive the main formula for the transverse distribution of the longitudinal fields $E_z$ and $H_z$ in the cross-section which includes a bunch. The method is based only on the assumption that a complete electromagnetic field in the cross-section of a bunch is equal to zero outside the vacuum channel. 

Let us consider the Maxwell system in CGS units inside the vacuum channel:
\begin{align}
\nabla\times\mathbf{E}&=-\frac{1}{c}\frac{\partial \mathbf{H}}{\partial t }, \nonumber \\ 
\nabla\times\mathbf{H}&=\frac{4\pi}{c}\mathbf{j}+\frac{1}{c}\frac{\partial \mathbf{E}}{\partial t }, \\ \nonumber
\nabla\cdot\mathbf{E}&=4\pi\rho, \\
\nabla\cdot\mathbf{H}&=0. \nonumber
\end{align}
We assume that the bunch is moving along the z-axis of the structure and the structure is longitudinally homogenous.  In the ultra-relativistic limit the current that produces the bunch and it's charge density could be written as:
\begin{align} 
\label{src}
j&=j_z= c \rho, \nonumber\\
\rho&=Q\delta(x-x_0)\delta(y-y_0)\delta(z-ct).
\end{align}
Let us introduce a new coordinate $\zeta=ct-z$ and split the Maxwell system into two parts:
\begin{align}
\label{Max1}
[\nabla\times\mathbf{E}]_\bot&=-\frac{\partial \mathbf{H_\bot}}{\partial \zeta }, \nonumber \\ 
[\nabla\times\mathbf{H}]_\bot&=\frac{\partial \mathbf{E_\bot}}{\partial \zeta };
\end{align}
and
\begin{align}
\label{Max2}
\nabla_\bot\times\mathbf{E}_\bot&=-\frac{\partial H_z}{\partial \zeta }, \nonumber \\ 
\nabla_\bot\times\mathbf{H}_\bot&=4\pi\rho+\frac{\partial E_z}{\partial \zeta }, \\ \nonumber
\nabla_\bot\cdot\mathbf{E}_\bot&=4\pi\rho+\frac{\partial E_z}{\partial \zeta }, \\
\nabla_\bot\cdot\mathbf{H}_\bot&=\frac{\partial H_z}{\partial \zeta }. \nonumber
\end{align}
Here $\bot$ -symbol indicates components orthogonal to the z-axis and $\nabla_\bot$ - is the nabla operator in the plane orthogonal to the z-axis.
From the equations \eqref{Max1} one can achieve:
\begin{align}
\label{K_R_c}
\frac{\partial H_z}{\partial x}&=\frac{\partial E_z}{\partial y},\\
\frac{\partial H_z}{\partial y}&=-\frac{\partial E_z}{\partial x}.  \nonumber
\end{align}
Equation \eqref{K_R_c} could be considered as the Cauchy-Riemann equation for some function defined in a complex plane. Let us consider the waveguide vacuum channel cross-section which includes the bunch as a complex plane. Let us rewrite \eqref{Max2} as:
\begin{align}
\label{Max3}
\nabla_\bot\cdot\mathbf{E}_\bot+i\nabla_\bot\times\mathbf{E}_\bot&=4\pi\rho+\frac{\partial E_z}{\partial \zeta }-i\frac{\partial H_z}{\partial \zeta }, \\
\nabla_\bot\cdot\mathbf{H}_\bot+i \nabla_\bot\times\mathbf{H}_\bot&=i\left(4\pi\rho+\frac{\partial E_z}{\partial \zeta }-i\frac{\partial H_z}{\partial \zeta }\right). \nonumber
\end{align}
We introduce complex functions:
\begin{align}
\label{c_fl}
e&=E_x+iE_y, \nonumber \\
h&=H_x+iH_y. 
\end{align}
We introduce an operator \cite{Morse}:
\begin{align}
\label{c_nab}
\nabla_c=\frac{\partial }{\partial x}+i\frac{\partial}{\partial y}.
\end{align}
One can see that
\begin{align}
\nabla_c^* e=\nabla_\bot \cdot \mathbf{E}_\bot+i\nabla_\bot\times\mathbf{E}_\bot, \nonumber \\
\nabla_c^* h=\nabla_\bot\cdot\mathbf{H}_\bot+i \nabla_\bot\times\mathbf{H}_\bot.
\end{align}
Here $*$ - symbol means the complex conjugation. 
On the other hand one can rewrite \eqref{c_nab} as:
\begin{align}
\label{diff}
\nabla_c&=2\frac{\partial}{\partial \chi^*}, \nonumber \\
\nabla_c^*&=2\frac{\partial}{\partial \chi}, 
\end{align}
here $\chi=x+iy$. 
From \eqref{K_R_c} one can treat function $L=\frac{\partial E_z}{\partial \zeta }-i\frac{\partial H_z}{\partial \zeta }$ as a plane vector field.  According to \cite{shabat}, as this field has no sources and vortexes, it could be found as a derivative of it's complex potential:
\begin{align}
\label{l_fl}
\frac{\partial{g_Z}}{\partial \chi}=\frac{\partial E_z}{\partial \zeta }-i\frac{\partial H_z}{\partial \zeta }.
\end{align}
By combining \eqref{Max3},\eqref{diff},\eqref{c_fl} and \eqref{l_fl} one can derive:
\begin{align}
\label{c_maxl}
2\frac{\partial e}{\partial \chi}&=4\pi \rho+\frac{\partial g_Z}{\partial \chi}. 
\end{align}
One can see that $h=ie$, thus we can consider only one equation for $e$. 

Let us consider a simply connected region $D$ with the arbitrary smooth boundary and corresponding complex plane $\omega$; with \eqref{src} equation \eqref{c_maxl} could be written as:
\begin{align}
\label{max_om}
2\frac{\partial e}{\partial \omega}&=4\pi Q\delta(x-x_0)\delta(y-y_0)\delta(-\zeta)+\frac{\partial g_Z}{\partial \omega}. 
\end{align}
According to the Riemann's theorem \cite{shabat} there exist a conformal mapping from the region $D$ to a circle and a corresponding $\chi$ - plane: $\chi=\chi(\omega)$ with $\chi(\omega_0)=0$ ($\omega_0=x_0+iy_0$). By introducing the substitution of variables in \eqref{max_om}, we arrive at:
  \begin{align}
\label{max_om2}
2\frac{\partial e}{\partial \chi}\frac{d\chi}{d\omega}&=4\pi Q \left | \frac{d\chi}{d\omega}\right |^2\delta(x')\delta(y')\delta(-\zeta)+\frac{\partial g_Z}{\partial \chi}\frac{d\chi}{d\omega}.
\end{align}   
Here $x'$ and $y'$ are defined as $\chi=x'+iy'$. Simplification of \eqref{max_om2} gives:
  \begin{align}
\label{max_omF}
2\frac{\partial e}{\partial \chi}&=4\pi Q \left ( \frac{d\chi}{d\omega}\right )^*\delta(x')\delta(y')\delta(-\zeta)+\frac{\partial g_Z}{\partial \chi}. 
\end{align}   
From \eqref{max_omF} we can see that the task of the derivation field $e$ in the $D$ region could be reduced to the task of the derivation $e$ in a circle with the source of intensity $4\pi Q\delta(-\zeta) \left.  \left (\frac{d\chi}{d\omega}\right )^*\right|_{\omega=\omega_0}$  placed in it's center.
Let us consider a circular contour $\Gamma$ in the $\chi$ - plane such that $|\chi|=r<a$ ($a$ -is the radius of a circle). According to \cite{shabat} with \eqref{max_omF}, a contour integral over $\Gamma$ could be written as:
\begin{align}
\label{o_int}
\oint\limits_\Gamma \left(e^*-\frac{g_Z^*}{2}\right)d\chi=4\pi i Q \delta(-\zeta) \left. \frac{d\chi}{d\omega}\right|_{\omega=\omega_0}.
\end{align}  
Now let us rely on the fact that in case of a cylindrical structure with the bunch driving along the center of a cylinder longitudinal components (and also their $\zeta$ derivatives) of the electric and magnetic field $E_z$,$H_z$ do not depend on the transverse coordinates. In this case, in $\chi$ plane from \eqref{l_fl} we have:
\begin{align}
\label{z_pot}
g_Z=C\chi,
\end{align}
where $C$ is a constant. Due to the symmetry of rotation in the $\chi$-plane the transverse field $e$ could be expressed as:
\begin{align}
\label{z_e}
e=\tilde{e}(|\chi|)\frac{\chi}{|\chi|}.
\end{align}
When placing \eqref{z_pot} and \eqref{z_e} into \eqref{o_int} and taking into account that $|e|$ is a constant on $\Gamma$ one can write:
\begin{align}
\label{o_int2}
\tilde{e}\oint\limits_{|\chi|=r} \frac{\chi^*}{|\chi|}d\chi-\frac{C^*}{2}\oint\limits_{|\chi|=r}\chi^*d\chi=4\pi i Q \delta(-\zeta) \left. \frac{d\chi}{d\omega}\right|_{\omega=\omega_0}.
\end{align}
Evaluation of the integrals on the left side of \eqref{o_int2} gives:
 \begin{align}
\label{e_mod}
\tilde{e}= \frac{2 Q\delta(-\zeta)}{r} \left. \frac{d\chi}{d\omega}\right|_{\omega=\omega_0}+\frac{C^*}{2}r,
\end{align}
multiplication of $\tilde{e}$ by $\frac{\chi}{|\chi|}$ and taking into account that $|\chi|=r$ gives:
 \begin{align}
\label{e_fl}
e= \frac{2 Q\delta(-\zeta)}{\chi^*} \left. \frac{d\chi}{d\omega}\right|_{\omega=\omega_0}+\frac{C^*\chi}{2}.
\end{align}
As we are considering the cross-section that includes a bunch and follow the idea in \cite{myPRL,myArxiv} we conclude that the field at the channel boundary vanish due to the phase speed of light in the surrounding medium being less than the speed of light in the channel, or we can say that the effective plane field $g_Z$ caused by the radiation prevents penetration of the free space field of the bunch into the medium. Taking this into account, one can write down a condition for the unknown constant $C$:
\begin{align}
\label{boundary_cond}
e(|\chi|=a)=0.
\end{align} 
Using this condition from \eqref{e_fl} we find:
\begin{align}
\label{c_for}
C=-\frac{4 Q\delta(-\zeta)}{a^2} \left. \left( \frac{d\chi}{d\omega}\right)^*\right|_{\omega=\omega_0}.
\end{align}
Thus:
\begin{align}
\label{e_fl2}
e= 2 Q\delta(-\zeta) \left. \frac{d\chi}{d\omega}\right|_{\omega=\omega_0} \left(\frac{1}{\chi^*}-\frac{\chi}{a^2} \right).
\end{align}
From \eqref{l_fl}, \eqref{z_e} and \eqref{c_for} we have:
\begin{align}
\label{final_ldir}
\frac{\partial E_z}{\partial \zeta }-i\frac{\partial H_z}{\partial \zeta }=-\frac{4 Q\delta(-\zeta)}{a^2}\frac{d\chi}{d\omega} \left. \left( \frac{d\chi}{d\omega}\right)^*\right|_{\omega=\omega_0}.
\end{align}
Finally at the point $\zeta=0$ (the cross-section that includes the bunch) we have:
\begin{align}
 E_z^0-i H_z^0=-\frac{2 Q}{a^2}\frac{d\chi}{d\omega} \left. \left( \frac{d\chi}{d\omega}\right)^*\right|_{\omega=\omega_0}.
\end{align}
if $\chi(\omega)=f(\omega)$ - function that gives the mapping of $D$ -region on a circle we can write result in a more compact form:
\begin{align}
\label{final_field}
 E_z^0=-\frac{2 Q}{a^2}\operatorname{Re}[f'(\omega)^*f'(\omega_0)], \\
 H_z^0=-\frac{2 Q}{a^2}\operatorname{Im}[f'(\omega)^*f'(\omega_0)] \nonumber. 
\end{align}
Formula \eqref{final_field} gives values of the longitudinal field components $E_z$ and $H_z$ in the cross-section of a  bunch which is moving along the vacuum channel of arbitrary shape. The formula is valid at the point of and away from the bunch and is the two-dimensional Green function by the transverse coordinates. By integrating  \eqref{final_field}  over $\omega_0$ with the transverse distribution of a bunch $\rho_\bot(\omega_0)$, one can achieve fields for the bunch distributed  in transverse coordinates.  

Another method for deriving formula \eqref{final_field} based  on the Poynting's theorem and more physical approach can be found in Appendix \ref{app:PT}.    

\section{Transverse part of the Lorentz force}
We consider a part of the Maxwell system \eqref{Max1} and write it down in the expanded form:
\begin{align}
\label{1}
\frac{\partial E_z}{\partial y}+\frac{\partial E_y}{\partial \zeta}=-\frac{\partial H_x}{\partial \zeta}, \\
\label{2}
-\frac{\partial E_x}{\partial \zeta}-\frac{\partial E_z}{\partial x}=-\frac{\partial H_y}{\partial \zeta}, \\
\label{3}
\frac{\partial H_z}{\partial y}+\frac{\partial H_y}{\partial \zeta}=\frac{\partial E_x}{\partial \zeta}, \\
\label{4}
-\frac{\partial H_x}{\partial \zeta}-\frac{\partial H_z}{\partial x}=\frac{\partial E_y}{\partial \zeta}. 
\end{align}
By combining  \eqref{1},\eqref{2} and \eqref{3},\eqref{4} we have:
\begin{align}
\label{for_exp}
2\frac{\partial}{\partial \zeta}\left(E_x-H_y\right)&=\frac{\partial H_z}{\partial y}-\frac{\partial E_z}{\partial x}, \nonumber \\
2\frac{\partial}{\partial \zeta}\left(E_y+H_x\right)&=-\frac{\partial H_z}{\partial x}-\frac{\partial E_z}{\partial y}.
\end{align}
Using the definition of the Lorentz force acting on a bunch we introduce a complex function $F_\bot$:
\begin{align}
\label{LF_def}
F_\bot=q\left[E_x-H_y+i(E_y+H_x)\right],
\end{align}  
where $q$ - is the charge of the test particle.
Let us rewrite \eqref{for_exp} with \eqref{LF_def} as:
\begin{align}
2\frac{\partial}{\partial \zeta}\left(F_x+iF_y\right)=q\left(\frac{\partial H_z}{\partial y}-\frac{\partial E_z}{\partial x}-i\left[\frac{\partial H_z}{\partial x}+\frac{\partial E_z}{\partial y}\right]\right).
\end{align}
In terms of the derivative \eqref{diff}  by  $\omega=x+iy$ \eqref{c_nab} the equation above simplifies to:
\begin{align}
\frac{\partial F^*_\bot}{\partial \zeta}=-q\frac{\partial}{\partial \omega}\left(E_z-iH_z\right).
\end{align}
In a vicinity of $\zeta=0$ with \eqref{final_ldir} we have 
\begin{align}
F_\bot=\frac{4 qQ \zeta}{a^2}\left(\frac{d^2\chi}{d\omega^2}\right)^* \left. \frac{d\chi}{d\omega}\right|_{\omega=\omega_0}.
\end{align}
If $\chi(\omega)=f(\omega)$ - function that gives the mapping of the $D$ -region on a circle we can write down the result in a more compact form:
\begin{align}
\label{final_Lor}
F_\bot=\frac{4 qQ \zeta}{a^2}f''(\omega)^* f'(\omega_0).
\end{align}
Formula \eqref{final_Lor} gives values of the transverse part of the Lorentz force acting on a bunch which is moving along the vacuum channel of arbitrary cross-section shape. By transverse coordinates, the formula is valid at the point of and away from the bunch and is the two-dimensional Green function. By the longitudinal coordinate, the formula is valid in a vicinity of  a point $\zeta=0$. By integrating  \eqref{final_Lor}  over $\omega_0$ with transverse distribution of a bunch $\rho_\bot(\omega_0)$ one can achieve the Lorentz force for the bunch distributed  in transverse coordinates. 

\section{Relativistic Gauss theorem} \label{sec:GT}
In this section we are going to show how one can derive the relativistic Gauss theorem introduced in \cite{myPRL,myArxiv} with the use of a formalism demonstrated in the Section \ref{sec:PF}.

Let us consider equation \eqref{max_om} and integrate it over the vacuum gap cross-section:
\begin{align}
2\int\limits_{S_{vac}}\frac{\partial e^*}{\partial \omega^*}dxdy&=4\pi Q\delta(-\zeta)+\int\limits_{S_{vac}}\frac{\partial g_Z^*}{\partial \omega^*}dxdy. 
\end{align}
Let us evaluate an integral on the left side according to the Green's theorem:
\begin{align}
\int\limits_{S_{vac}}\frac{\partial e^*}{\partial \omega^*}dxdy=-i/2\oint\limits_{\Gamma'} e^* d\omega,
\end{align}
here $\Gamma'$ - is the boundary of the vacuum channel region $S_{vac}$. 

Taking into account condition \eqref{boundary_cond} ($e$=0 on $\Gamma'$)  with \eqref{l_fl} we immediately have: 
\begin{align}
\frac{\partial}{\partial \zeta }\left(\int\limits_{S_{vac}}E_zdS+i\int\limits_{S_{vac}} H_zdS\right)=-4\pi Q\delta(-\zeta).
\end{align}
In a vicinity of the the point $\zeta=0$ we have:
\begin{align}
\label{gauss_theor}
\int\limits_{S_{vac}}E_zdS+i\int\limits_{S_{vac}} H_zdS=-4\pi Q\theta(-\zeta),
\end{align}
here $\theta(-\zeta)$ is a Heaviside theta-function.

Expression \eqref{gauss_theor} is a slightly generalized form of the relativistic Gauss theorem formulated in \cite{myPRL,myArxiv}. However, equation \eqref{gauss_theor} states that an integral over the vacuum channel from the magnetic field $H_z$ is zero in the cross-section of the bunch and in a vicinity of the the point $\zeta=0$. This result was not mentioned in \cite{myPRL}. 
\section{Conclusion}
In this paper we have derived general formula for the longitudinal field components $E_z$ and $H_z$ \eqref{final_field} in the cross-section of a bunch and a vicinity of the point $\zeta=0$. The formula expresses the field though the derivative of the function that conformally remaps the channel of the form one is interested in on a circle . It is worth to mention that the mapping function itself does not appear in the formula and one needs to know only it's derivative. It simplifies the calculations even more  as, for example, in case of the Christoffel-Schwarz integral fields expressed in terms of the function under the Christoffel-Schwarz integral. 

Now, if one take the point of the bunch (which corresponds to the $\omega=\omega_0$ in \eqref{final_field}), we arrive at the result of the paper \cite{myPRL}:
\begin{align}
E_z(0)=-\frac{2Q}{a^2}|f'(\omega_0)|^2=-\frac{2Q}{a^2}|J|,
\end{align}     
here $|J|$ is the determinants of the Jacobi matrix at the point of the bunch.

We also derived a formula for the transverse Lorentz force \eqref{final_Lor}, which could be used for the bunches  with non-point transverse distribution.

\renewcommand{\abstractname}{Acknowledgements}
\begin{abstract}
 The author is grateful to A.D. Kanareykin for useful discussions and suggestions.
\end{abstract}

\appendix
\section{}
\subsection{Derivation based on the Poynting's theorem} \label{app:PT}
\smallskip
Let us start from the Poynting's theorem
\begin{align}
\label{PTH}
	\frac{\partial u}{\partial t}+\mathbf{\nabla}\cdot \mathbf{S}=-\mathbf{j}\mathbf{E}.
\end{align}
Here
\begin{align}
	\mathbf{S}=\frac{c}{4\pi}[\mathbf{E}\times\mathbf{H}],\\
\nonumber	 
	u=\frac{1}{8\pi}\left(\mathbf{E}\mathbf{D}+\mathbf{B}\mathbf{H}\right).
\end{align}
Using the fact that the square of the transverse part of the Lorentz force divided by the charge $\mathbf{f}_\bot=\mathbf{F}_\bot/Q$  inside the vacuum channel could be expressed as:
\begin{align}
{{f}_{\bot }}^{2}={{E}_{\bot }}^{2}+{{H}_{\bot }}^{2}-2{{S}_{z}},
\end{align}
and introducing a new coordinate  $\zeta=ct-z$ one can rewrite \eqref{PTH} for the vacuum channel as follows:
\begin{align}
\label{inter}
\frac{\partial (E_{z}^{2}+H_{z}^{2})}{2\partial \zeta }+\frac{\partial f_{\bot }^{2}}{2\partial \zeta }+\frac{4\pi }{c}{{\nabla }_{\bot }}\cdot {{\mathbf{S}}_{\bot }}=-\frac{4\pi }{c}\mathbf{jE}.
\end{align}
Taking into account the fact that the bunch is moving along the z -axis and 
\begin{align}
	j=Qc\delta (x-{{x}_{0}})\delta (y-{{y}_{0}})\delta (-\zeta ),
\end{align}
equation \eqref{inter} takes the following form:
\begin{align}
\label{vch_1}
\frac{\partial (E_{z}^{2}+H_{z}^{2})}{2\partial \zeta }+\frac{\partial f_{\bot }^{2}}{2\partial \zeta }+\frac{4\pi }{c}{{\nabla }_{\bot }}\cdot {{\mathbf{S}}_{\bot }}=-4\pi Q{{E}_{z}}\delta (x-{{x}_{0}})\delta (y-{{y}_{0}})\delta (-\zeta ).
\end{align}
Now let us integrate \eqref{vch_1} over the cross-section which includes the bunch:
\begin{align}	
\int{\left[ \frac{\partial (E_{z}^{2}+H_{z}^{2})}{2\partial \zeta }+\frac{\partial f_{\bot }^{2}}{2\partial \zeta } \right]}dS+\frac{4\pi }{c}\int{{{\mathbf{S}}_{\bot }}\mathbf{n}dl}=\\ \nonumber 
=-4\pi Q\delta (-\zeta )\int{{{E}_{z}}\delta (x-{{x}_{0}})\delta (y-{{y}_{0}})dS}.
\end{align}
Taking into account that energy flux in the cross-section of the bunch  is collinear to the beam direction, the orthogonal component of the Ponynting vector should be equal to zero as a consequence of the transverse Lorentz force in this cross-section being zero. By integrating over $\zeta$  from $-\infty$ to zero, one can derive:
\begin{align}
\label{main_balance}
	\int{\left[ E_{z}^{2}+H_{z}^{2} \right]}dS=-4\pi Q\int{{{E}_{z}}\delta (x-{{x}_{0}})\delta (y-{{y}_{0}})dS}.
\end{align}
Here integrals are taken over the vacuum gap cross-section.
Let us also take into account the consequence of the Maxwell system: the equation for longitudinal components of the electric and magnetic field in an ultra-relativistic case inside the channel:
\begin{align}
\label{tr_lap}
	\Delta_{\bot }E_{z}=0, \\ \nonumber
	\Delta_{\bot}H_{z}=0.
\end{align}
Here  $\Delta_\bot$ - is the transverse part of the Laplace operator. From equations \eqref{tr_lap} one can conclude that $E_z$   and $H_z$ are harmonic functions inside the vacuum channel and fulfil the Cauchy-Riemann equations \eqref{K_R_c}. This means that one can introduce a complex analytical function inside the vacuum channel
\begin{align}
	w={{E}_{z}}-i{{H}_{z}}.
\end{align}
The modulus square of this function according to \eqref{main_balance} will give energy density inside the cross-section:
\begin{align}
\label{ENG}
	|w|^2=E_z^2+H_z^2.
\end{align}

Let us consider the case of a cylindrical structure when the bunch is driving along the axis of the cylinder  \cite{myPRL,myArxiv}:
\begin{align}
\label{PRL_f}
	E_z=-\frac{2q}{a^2}.
\end{align}
Let us consider a conformal mapping of the arbitrary cross-section ($\omega$-plane )on the circle  ($\chi$-plane):
\begin{align}
	\chi=f(\omega).
\end{align}
In this case formula \eqref{main_balance} in the coordinates of the $\chi$ - plane could be written as:
\begin{align}	
	\int{\left[ E_{z}^{2}+H_{z}^{2} \right]}{{\left| {f}'(\omega ) \right|}^{2}}d{{S}^{\omega }}=4\pi \int{{{E}_{z}}}\delta \left[ r({{x}_{1}},{{x}_{2}}) \right]{{\left| {f}'(\omega ) \right|}^{2}}d{{S}^{\omega }}.
\end{align}
An integral on the right side is equal to
\begin{align} 
	4\pi Q\int{{{E}_{z}}}\delta \left[ r({{x}_{1}},{{x}_{2}}) \right]{{\left| {f}'(\omega ) \right|}^{2}}d{{S}^{\omega }}=4\pi Q{{E}_{z}}{{\left| {f}'({{\omega }_{0}}) \right|}^{2}}.
\end{align}
Here  $\omega_0$ - is the point in the $\omega$  - plane that corresponds to the center of the circle in the  $\chi$ - plane. On the other hand in the $\chi$ - plane we have:
\begin{align}
	4\pi Q\int{{{E}_{z}}}\delta \left[ r({{x}_{1}},{{x}_{2}}) \right]d{{S}^{\chi }}=4\pi Q{{E}_{z}}=\int{\left[ E_{z}^{2}+H_{z}^{2} \right]}d{{S}^{\chi }},
\end{align}
thus 
\begin{align}
	\int{\left[ \tilde{E}_{z}^{2}+\tilde{H}_{z}^{2} \right]}d{{S}^{\omega }}={{\left| {f}'({{\omega }_{0}}) \right|}^{2}}\int{\left[ E_{z}^{2}+H_{z}^{2} \right]}{{\left| {f}'(\omega ) \right|}^{2}}d{{S}^{\omega }}.
\end{align}
Taking into account that  $E_z$ field in the $\chi$ - plane is defined by \eqref{PRL_f} and $H_z=0$ , one can write:
\begin{align}
	\int{\left[ \tilde{E}_{z}^{2}+\tilde{H}_{z}^{2} \right]}d{{S}^{\omega }}=\frac{4{{Q}^{2}}}{{{a}^{4}}}{{\left| {f}'({{\omega }_{0}}) \right|}^{2}}\int{{{\left| {f}'(\omega ) \right|}^{2}}d{{S}^{\omega }}}.
\end{align}
From \eqref{ENG} we conclude that
\begin{align}
\label{w_mod}
	|w|=-\frac{2Q}{{{a}^{2}}} |{f}'({{\omega }_{0}})||{f}'(\omega)|.
\end{align}
From \cite{myPRL} we know that
\begin{align}
w(\omega_0)=E_z=-\frac{2Q}{{{a}^{2}}} |{f}'({{\omega }_{0}})|^2.
\end{align}
This means that there are only two possible variants of the function $w$:
\begin{align}
\label{w1}
w=-\frac{2Q}{{{a}^{2}}} {f}'({{\omega }_{0}})^*{f}'(\omega), 
\end{align}
or
\begin{align}
\label{w2}
w=-\frac{2Q}{{{a}^{2}}} {f}'({{\omega }_{0}}){f}'(\omega)^*.
\end{align}
Equations \eqref{w1} and \eqref{w2} differ by the sign of a longitudinal component of the magnetic field. Taking into account \eqref{K_R_c}  and the fact that $f(\omega)$ is an analytic function and satisfies the Cauchy-Riemann equations, we conclude:
\begin{align}
\label{f_res}
	{{E}_{z}}=-\frac{2Q}{{{a}^{2}}}\operatorname{Re}\left[{f}'(\omega )^* {f}'({{\omega }_{0}}) \right], \\
	{{H}_{z}}=-\frac{2Q}{{{a}^{2}}}\operatorname{Im}\left[{f}'(\omega )^* {f}'({{\omega }_{0}}) \right].
\end{align}
One can see that the result \eqref{f_res} is equal to the result \eqref{final_field} of a Section \ref{sec:PF}.
\section{}
In this appendix we present results of the transverse structure calculations of the longitudinal electric field $E_z$ in the cross-section of a bunch for cylindrical (\ref{app:cyl}) and planar (\ref{app:planar}) structures. The results achieved using formula \eqref{final_field}  are compared with the calculations using mode decomposition method \cite{NG_cyl,Altmark_cyl,mySTAB} and are in full agreement. In Section \ref{app:rectan} we show how the suggested method could be combined with the mirror charges technique to derive a field in case metal walls are present in the cross-section of a bunch. Rectangular waveguide with side metal walls is considered as an example  and formula for the longitudinal electric field $E_z$ at the point of a bunch in case the bunch is driving in the center of the structure derived.

\subsection{Transverse distribution of the $E_z$ field in a cylindrical waveguide} \label{app:cyl}
\begin{figure}
\centering
\includegraphics[scale=0.7]{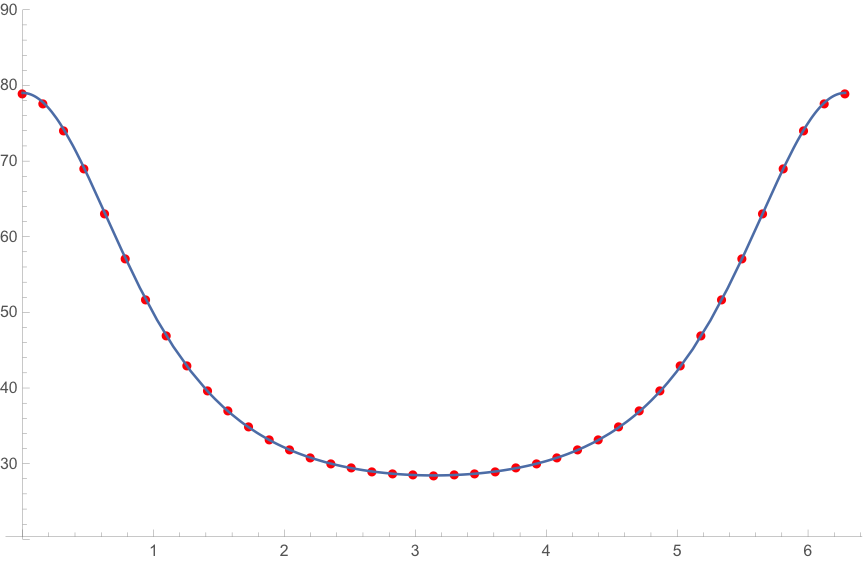}
\caption{ Angular dependence of $E_z$ in case of the displaced bunch and $r=r_0$. Red dots are for the direct simulation using the mode decomposition method \cite{NG_cyl,Altmark_cyl} and a solid line is for the formula \eqref{cyl_res}.}
\label{fig:1}
\end{figure}
\begin{figure}
\centering
\includegraphics[scale=0.7]{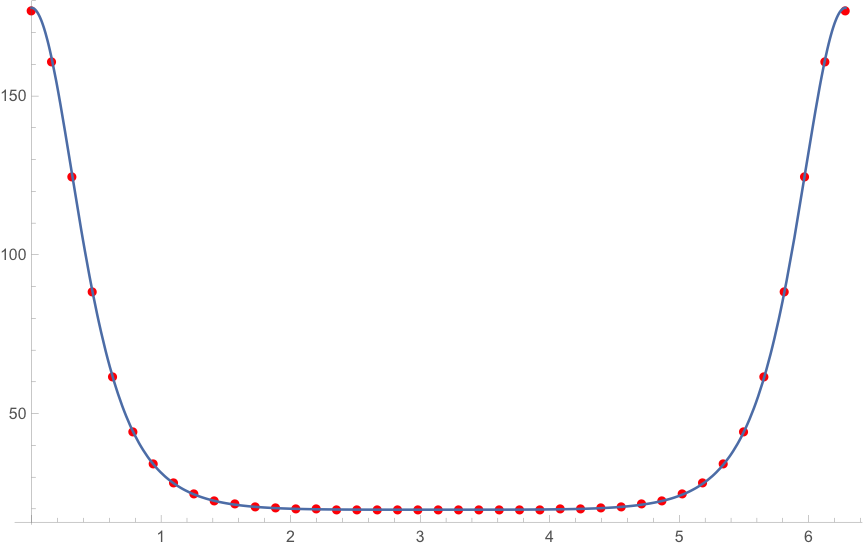}
\caption{Angular dependence of $E_z$ in case of the displaced bunch and $r=a$ (on the channel boundary). Red dots are for the direct simulation using the mode decomposition method \cite{NG_cyl,Altmark_cyl} and a solid line is for the formula \eqref{cyl_res}.}
\label{fig:2}
\end {figure}

\begin {figure}
\centering
\includegraphics[scale=0.7]{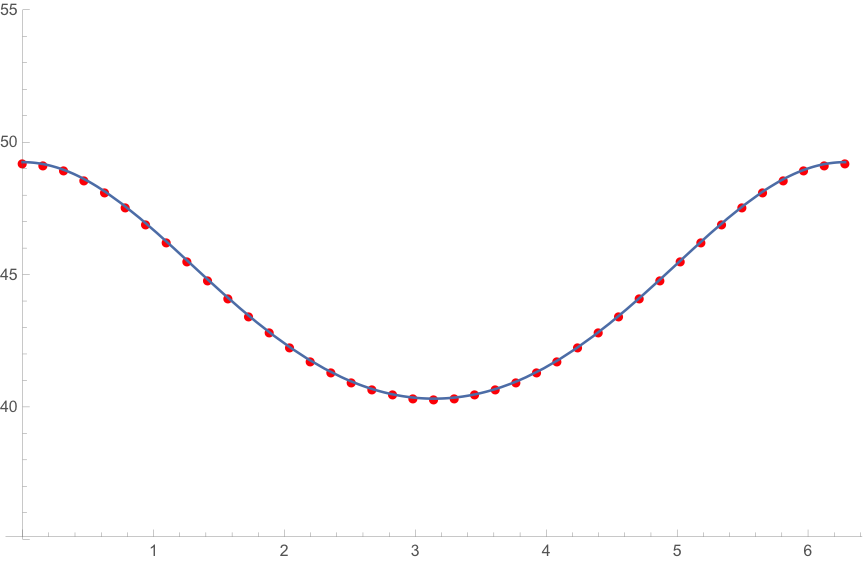}
\caption{Angular dependence of $E_z$ in case of the displaced bunch and $r=a/10$. Red dots are for the direct simulation using the mode decomposition method \cite{NG_cyl,Altmark_cyl} and a solid line is for the formula \eqref{cyl_res}.}
\label{fig:3}
\end {figure}
Conformal mapping of a circle with the radius $a$ on to a circle with radius $a$  such that the point $\omega_0=r_0$  corresponds to the center of the second circle is given by
\begin{align}
\label{Cyl_map}
	f(\omega)=a^2\frac{\omega-r_0}{a^2-\omega r_0}.
\end{align}

Derivative is given by:
\begin{align}
	f'(\omega)=\frac{a^2(a^2-r_0^2)}{(a^2-r_0\omega)^2}.
\end{align}
\smallskip
 \begin {table}[h!]
 \begin{center}
\caption{Parameters for the cylindrical structure.}
\smallskip
\label{tab:Table1}
\begin {tabular*}{0.75\textwidth}{@{\extracolsep{\fill}} c c c c c}
\toprule $a$&$b$& $\varepsilon$&$r_0$&$Q$ \\
\midrule
~~0.3~cm&0.32~cm & 5.7& 0.15~cm&1~nC~~ \\
\bottomrule
\end {tabular*}
\normalsize
\end {center}
\end{table}
Assuming $\omega=r\mathrm{exp[i\phi]}$  one can achieve:
\begin{align}
	f'(\omega_0)=\frac{a^2}{a^2-r_0^2},
\end{align}
and
\begin{align}
	\operatorname{Re}[{f}'(\omega )^*]=\frac{{{a}^{2}}({{a}^{2}}-r_{0}^{2})({{a}^{4}}-2{{a}^{2}}r{{r}_{0}}\cos (\phi )+{{r}^{2}}r_{0}^{2}\cos (2\phi )}{{{({{a}^{4}}+{{r}^{2}}r_{0}^{2}-2{{a}^{2}}r{{r}_{0}}\cos (\phi ))}^{2}}}.
\end{align}
From \eqref{f_res} we have:
\begin{align}
\label{cyl_res}
	{{E}_{z}}=-\frac{2Q}{{{a}^{2}}}\frac{{{a}^{8}}-2{{a}^{6}}r{{r}_{0}}\cos (\phi )+{{r}^{2}}r_{0}^{2}{{a}^{4}}\cos (2\phi )}{{{({{a}^{4}}+{{r}^{2}}r_{0}^{2}-2{{a}^{2}}r{{r}_{0}}\cos (\phi ))}^{2}}}.
\end{align}

Table.\ref{tab:Table1} contains a list of parameters for the cylindrical structure. Results are depicted in Figure.\ref{fig:1}, Figure.\ref{fig:2} and Figure.\ref{fig:3}. Figures show angular dependence of the longitudinal electric field $E_z$ for three different values of the test charge radial coordinates. We see full agreement with the mode decomposition method \cite{Altmark_cyl}.

\subsection{Transverse distribution of the $E_z$ field in a planar waveguide} \label{app:planar}

Conformal mapping of a strip with the half-high $a$  on a circle with the radius $a$  such that the point $\omega_0=0$   corresponds to the center of the circle is given by:
\begin{align}
	f(\omega )=a\tan \left[ \frac{\pi \omega }{4a} \right].
\end{align}
Derivative is given by:
\begin{align}
	{f}'(\omega )=\frac{\pi }{4}{{\left( \sec \left[ \frac{\pi \omega }{4a} \right] \right)}^{2}}.
\end{align}

\begin {figure}
 \centering
\includegraphics[scale=0.7]{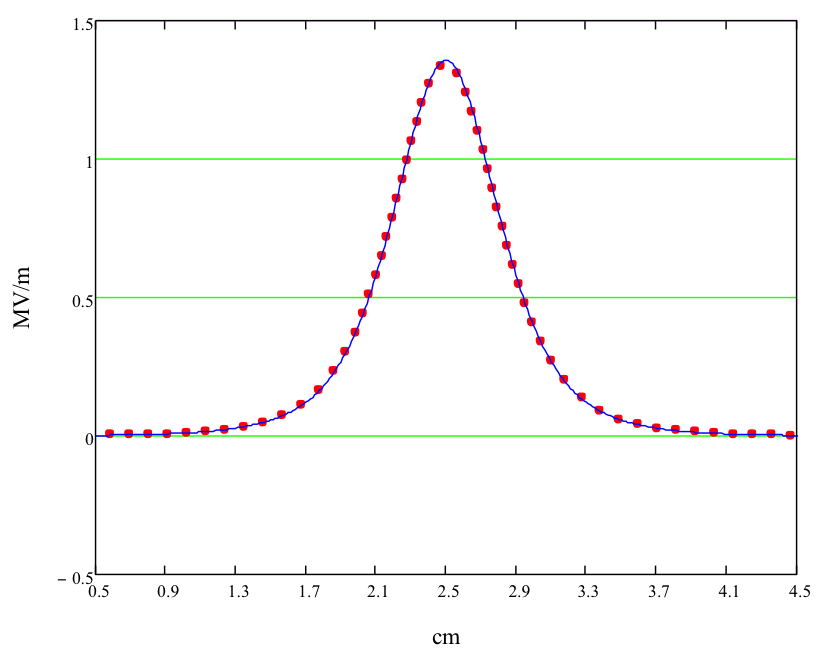}
\caption{ Y-dependence of $E_z$ in case the test charge is located at $x=0.4$ cm from the center. Red dots are for the direct simulation using the mode decomposition method \cite{mySTAB} and a solid line is for the formula \eqref{pl_res}.}
\label{fig:4}
\end {figure}

 \begin {figure}
 \centering
\includegraphics[scale=0.7]{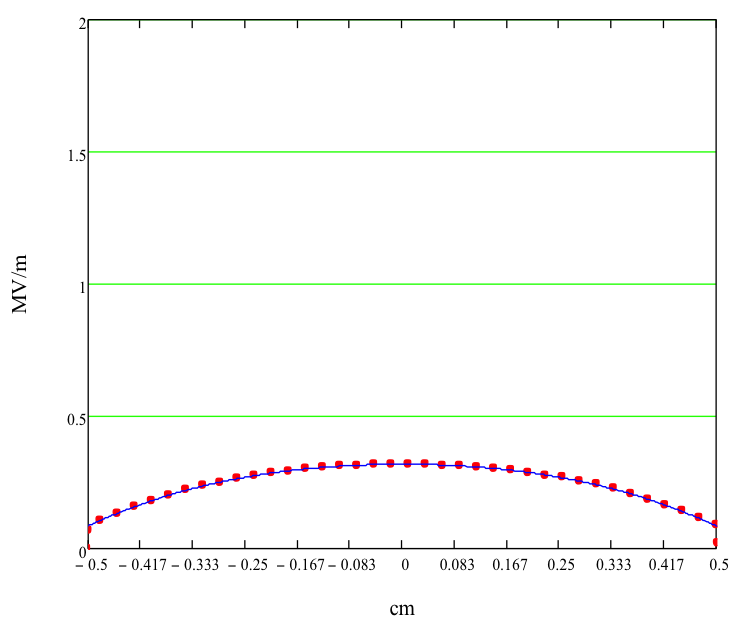}\\
\caption{X-dependence of $E_z$ in case the test charge is located at $y=0.7$ cm from the center. Red dots are for the direct simulation using the mode decomposition method \cite{mySTAB} and a solid line is for the formula \eqref{pl_res}.}
\label{fig:5}
\end {figure}

\smallskip
 \begin {table}[h!]
 \begin{center}
\caption{Parameters for the planar structure.}
\smallskip
\label{tab:Table2}
\begin {tabular*}{0.75\textwidth}{@{\extracolsep{\fill}} c c c c c}
\toprule $a$&$b$& $\varepsilon$&$r_0$&$Q$ \\
\midrule
~~0.5~cm&0.502~cm & 5.7& 0~cm&1~nC~~ \\
\bottomrule
\end {tabular*}
\normalsize
\end {center}
\end{table}
Assuming $\omega=x+iy$,  one can achieve:
\begin{align}
	{f}'({{\omega }_{0}})=\frac{\pi }{4}.
\end{align}
and
\begin{align}
	\operatorname{Re}[{f}'(\omega )^*]=\frac{\pi }{4}\frac{1+\cos \left[ \frac{\pi x}{2a} \right]\cosh \left[ \frac{\pi y}{2a} \right]}{{{\left( \cos \left[ \frac{\pi x}{2a} \right]+\cosh \left[ \frac{\pi y}{2a} \right] \right)}^{2}}}.
\end{align}

From \eqref{f_res} we have:
\begin{align}
\label{pl_res}
	{{E}_{z}}=-\frac{2Q}{{{a}^{2}}}\frac{{{\pi }^{2}}}{8}\frac{1+\cos \left[ \frac{\pi x}{2a} \right]\cosh \left[ \frac{\pi y}{2a} \right]}{{{\left( \cos \left[ \frac{\pi x}{2a} \right]+\cosh \left[ \frac{\pi y}{2a} \right] \right)}^{2}}}.
\end{align}

Table.\ref{tab:Table2} contains a list of parameters for the planar structure. Results are depicted on Figure.\ref{fig:4} and Figure.\ref{fig:5}. Figures show coordinate dependence of the longitudinal electric field $E_z$ on the $x$ and $y$ coordinates. We see full agreement with the mode decomposition method \cite{mySTAB}.

\subsection{Longitudinal electric field $E_z$ on a bunch in a rectangular structure with side metal walls} \label{app:rectan}

We consider a rectangular structure which differs from the planar by two parallel perfect conducting walls. In case the field distribution in the vacuum gap is known for the planar case \eqref{pl_res}, contribution of the metal walls could be included by introducing mirror charges. If the field for the planar case is $E_z^{pl}(x,y)$ and metal walls are placed at $y=\pm w/2$, a full field could be found as a superposition of the real charge field and imaginary charges field. In case the charge is placed in the centre of the 
structure, a full field could be found as:
\begin{align}
E_z^{rec}(x,y)=E_z^{pl}(x,y)+\sum\limits_{n=1}^\infty(-1)^n E_z^{pl}(x,y-nw)+\sum\limits_{n=1}^\infty(-1)^n E_z^{pl}(x,y+nw).
\end{align}

The logitudinal electric field $E_z$ at the point of a bunch could be found as:
\begin{align}
\label{loss_rec}
E_z^{rec}(0,0)=E_z^{pl}(0,0)+2\sum\limits_{n=1}^\infty(-1)^n E_z^{pl}(0,nw).
\end{align}
\begin {figure}
 \centering
\includegraphics[scale=0.65]{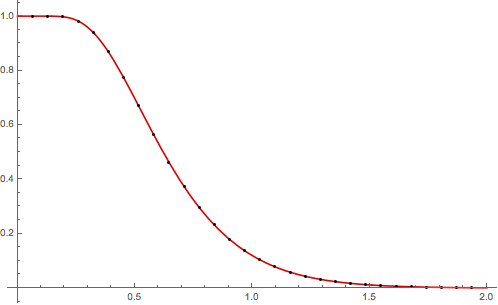}\\
\caption{Normalized longitudinal electric field $E_z$ on a bunch in the rectangular waveguide; dependence from the waveguide relative width $a/w$. Red line - dependence of formula \eqref{loss_rec_cnt} on $a/w$ normalized by $-\frac{2Q}{a^2}\frac{\pi^2}{16}$, black dots - same dependence achieved using the mode decomposition method \cite{mySTAB} and also normalized by $-\frac{2Q}{a^2}\frac{\pi^2}{16}$.}
\label{fig:6}
\end {figure}
By placing \eqref{pl_res} into \eqref{loss_rec}, one can achieve:
\begin{align}
\label{loss_rec_cnt}
E_z^{rec}(0,0)=-\frac{2Q}{a^2}\frac{\pi^2}{16}\left(1+4\sum\limits_{n=1}^\infty \frac{(-1)^n}{1+\mathrm{cosh}\left( \frac{ \pi nw}{2a}\right)} \right).
\end{align}
Figure.\ref{fig:6} shows dependence of the normalized longitudinal electric field $E_z$ at the point of the bunch on a relative width of the waveguide - $a/w$. One can see full agreement with the mode decomposition method \cite{mySTAB}. 

\newpage

\bibliographystyle{ieeetr}

\begin{thebibliography}{15}
\bibitem{ILC} \href{http://linearcollider.org}{ILC Technical Design Report}.
\bibitem{CLIC} M. Aicheler. A Multi-TeV Linear Collider Based on CLIC Technology: \href{http://cds.cern.ch/record/1500095?ln=en}{CLIC Conceptual Design Report}.
\bibitem{Bost}C Bostedt et al. J. Phys. B $\mathbf{46}$, 164003, (2013).
\bibitem{Bane1}K.F.Bane. Wakefields of sub-picosecond electron bunches. SLAC-pub-11829, (2006). 
\bibitem{Bane2}K.L.F. Bane, G. Stupakov. Nucl. Instr. Meth. Phys. Res. A $\mathbf{690}$, p.106, (2012).
\bibitem{Novo}A. Novokhatski, A. Mosnier, Proc.of PAC 1997, Vancouver, 1997, p. 1661, (1997)
\bibitem{Bane3}K. Bane and G. Stupakov, Phys. Rev. ST-Accel. Beams $\mathbf{6}$, 024401 (2003).
\bibitem{myPRL}  S.S. Baturin and A.D. Kanareykin, Phys. Rev. Lett $\mathbf{113}$, 214801 (2014).
\bibitem{myArxiv} S.S. Baturin and A.D. Kanareykin, \href{http://arxiv.org/abs/1409.0209}{arxiv:1409.0209} (2014).
\bibitem{myArxiv_leont} S.S. Baturin and A.D. Kanareykin, \href{http://arxiv.org/abs/1308.6228}{arxiv:1308.6228} (2014).
\bibitem{Morse} P.M. Morse, H. Feshbach, \textit{Methods of Theoretical Physics. Part I.}, Chapter 4, P. 330, McGraw-Hill Science/Engineering/Math (1953).
\bibitem{shabat} M.A. Lavrentiev, B.V. Shabat, \textit{Methods of Complex Function Theory.}, Nauka, Moskow, (1987).
\bibitem{NG_cyl}  K.-Y. Ng, Phys. Rev. D, $\mathbf{42}$, p. 1819 (1990).
\bibitem{Altmark_cyl} A.M. Altmark, A.D. Kanareykin, Journal of Physics Conf. Series, $\mathbf{357}$, 012001 (2012).
\bibitem{mySTAB} S.S. Baturin, I.L. Sheinman, A.M. Altmark, A.D. Kanareykin,  Phys. Rev. ST-AB., $\mathbf{16}$, 051302 (2013).
\end{thebibliography}

\end{document}